\documentclass[11pt,a4paper]{article}
\pdfoutput=1
\usepackage{jheppub}

\usepackage{amsmath}
\usepackage{amssymb}
\usepackage{graphicx}

\usepackage{color}

\newcommand{\be}{\begin{equation}}
\newcommand{\ee}{\end{equation}}
\newcommand{\bea}{\begin{eqnarray}}
\newcommand{\eea}{\end{eqnarray}}
\newcommand{\nn}{\nonumber}

\newcommand{\sign}{\text{sign}}

\title{Holographic Thermalization  from Kerr-AdS}

\author[a]{Irina Aref'eva,}
\author[b]{Andrey Bagrov,}
\author[c]{Alexey S. Koshelev}
\affiliation[a]{Steklov Mathematical Institute, RAS, Gubkin str. 8, 119991
Moscow, Russia}
\affiliation[b]{Lorentz Institute for Theoretical Physics, Leiden
University,
Niels Bohrweg 2, 2333CA Leiden,
The Netherlands}
\affiliation[c]{Theoretische Natuurkunde, Vrije Universiteit Brussel and The
International Solvay Institutes, Pleinlaan 2, B-1050 Brussels,
Belgium}
\emailAdd{arefeva@mi.ras.ru}
\emailAdd{bagrov@lorentz.leidenuniv.nl}
\emailAdd{alexey.koshelev@vub.ac.be}

\abstract{We study thermalization of a strongly coupled theory
holographically dual to a thin shell of null dust with non-zero
angular momentum collapsing to Kerr-AdS.
 We calculate thermalization time for two point correlation functions. It
happens that in the 3-dimensional case  the thermalization time is
just proportional to the distance between points where
the correlator is evaluated. This is a very surprising and rather
unexpected generalization of the same relation in the case of zero
momentum.}

\keywords{Holography and quark-gluon plasmas, gauge-gravity
correspondence}
\begin{document}
\maketitle

\section{Introduction}
Among many questions one of the most intriguing problem in the
physics of heavy ion collisions (HIC) is understanding of the
short thermalization time, less than $1$ fm/c. At later times a
local thermodynamical equilibrium settles, and a hydrodynamic
description of the quark-gluon plasma (QGP) is possible. The main
difficulty is that describing the thermalization processes one has
to deal with time evolution of strongly coupled systems.

A powerful approach to such
problems was developed on
the basis of
the holographic duality between a strongly coupled
quantum field theory in $d$-dimensional Minkowski space and the classical
gravity in $d+1$-dimensional anti-de Sitter space (AdS)
\cite{Maldacena,Klebanov,Witten}
during the last few years. In particular, there is a considerable
progress in
 the holographic description of the equilibrium  QGP  \cite{1101.0618}.
This technique can also be applied to non-equilibrium quantum
systems. Within this framework the thermalization is described as
a process of black hole (BH) formation in AdS.

The gravitational collapse of matter injected into Minkowski
space-time and the formation of event horizon are old problems in
general relativity. The same question might be asked in AdS,
 and one can analyze what deformations of AdS metric end up with the BH
formation.
 In the context of the application of  AdS/CFT to HIC it is interesting to
  consider deformations which can be interpreted as a dual to
  the initial state of relativistic heavy ions.

Various models have been proposed in the literature:
\begin{itemize}
\item
colliding gravitational shock waves \cite{Gubser:2008pc}-\cite{Taliotis:2012sx},
\item colliding shell of matter with vanishing rest mass (``null dust''),
the so called AdS-Vaidya models \cite{Keski}-\cite{1212.6066},
\item
sudden near-boundary perturbations of the metric propagating
into the bulk \cite{Chesler:2009cy,Chesler:2008hg,Bhattacharyya:2009uu},
\item
coupled quenches
\cite{Das:2010yw,Hashimoto:2010wv}.
\end{itemize}
As stated above, an explanation of the
short formation time of QGP (short thermalization time) in HIC
using AdS/CFT techniques is an important and an ambitious goal.
For a review we refer the reader to \cite{IA-poc,Gubser-review}, and refs.
therein.

Currently the AdS-Vaidya metric
\cite{Keski}-\cite{1212.6066} seems to be the most tractable model
to answer this question, and we focus on it in this paper. This is a very
instructive holographic toy model that captures
many important features of quantum fields thermalization. In this
model a thin collapsing null dust shell in $AdS$ is considered.
The final black hole state is dual to the thermalized equilibrium boundary
QFT.

There is a possibility to write analytical, but implicit formula
for the thermalization time of two-point correlators for a wide
class of backgrounds, like  Reissner-Nordstr\"om-AdS and Lifshitz-AdS
\cite{Keranen:2011xs}-\cite{Tarrio:2011de}. In $2+1$ dimensional
case the corresponding formula can be written explicitly. Numerical analysis of
the thermalization time in higher dimensions indicates that it is
nothing more but the causal bound on the possible space of an after
quench evolution.

Results in \cite{Garfinkle:2011hm,Garfinkle:2011tc,Wu:2012rib} also show very
rapid  BH
formation, close to the causal bound for a very
wide range of BH masses. In the BTZ-Vaidya metric in $AdS_3/CFT_2$
case, we can write an explicit formula for the thermalization time
 \cite{Balasubramanian:2011ur,Lopez,Callan:2012ip}.
 Note that in those papers only the BTZ-Vaidya metric at
zero angular momentum has been considered.

In the present work we extend the analysis of
\cite{Balasubramanian:2011ur,Lopez,Callan:2012ip} to the case of
rotating null dust shell collapsing into AdS-Kerr-BTZ BH
and show that the thermalization time of a non-equal time two point
correlators in a field theory dual to this gravitational
background can be calculated exactly, and, rather unexpectedly, a presence
of the second event horizon and a non-zero angular momentum does not
affect it at least in the (2+1)-dimensional case. To
study the same question in higher dimensions one is required
to perform a numerical analysis.

Another branch of the contemporary research on AdS/CFT that our
considerations might be relevant for is the holographic
information theory. One of the cornerstones of modern applications
of quantum field theory to solid state physics is the concept of
the entanglement entropy \cite{Srednicki:1993im}.
{It seems to be intimately related with ground
states of strongly interacting field theories, and is widely used
in theoretical many-body physics as a classifier of both quantum
critical \cite{Huijse} and topological phases \cite{Wen}.}

This quantity is very hard to calculate using standard field
theoretical techniques. Holographic solution to this problem has
been proposed in \cite{Ryu-Takayanagi} in a form of conjecture,
recently proven in \cite{Maldacena-Lewkowycz}. It says that the entanglement
entropy of a region ${\cal A}$ on the boundary is proportional to
area of minimal surface in the bulk that has a common boundary with
${\cal A}$.

The advantage of a low dimensional holography is that in contrary to
a higher dimensional case both two-point correlators and entropy are
encoded in the same objects in the bulk, - in space-like
geodesics. Performing the geodesic analysis we are testing
the thermalization properties of observables of both kinds
at the same time. In this regard our results on thermalization time
are universal (even though we do not discuss here subtle aspects of
the entanglement entropy phenomenology).

The paper is organized as follows. In section 2 we briefly recall
the BTZ solution \cite{btz}. Sections 3 and 4 form the central
part of the paper. In 3.1 and 3.2 we provide a detailed derivation
of space-like geodesics in this space-time (in different
parametrizations). In 3.3 we analyze properties of geodesics of a
particular kind, which start and end up at the conformal boundary
and therefore related to two-point correlation functions in the
boundary field theory. In Section 4 we consider evolution of these
geodesics in the background of a collapsing null shell and
calculate holographically the thermalization time of related
boundary correlation functions.

\section{$D=3$ Kerr-BTZ}

The original BTZ formula is
\begin{eqnarray}\label{BTZ-o}
ds^2 &=& -(-M+\frac{r^2}{l^2})d t^2 + \frac{dr^2}{-M +
\frac{r^2}{l^2} + \frac{a^2}{r^2}} -2adtd\phi +
r^2d\phi^2.\end{eqnarray} We introduce for the sequel
\begin{equation}
{\cal
K}={r}^{2}M{l}^{2}-{r}^{4}-{a}^{2}{l}^{2}=-(r^2-\beta_1)(r^2-\beta_2),
\end{equation}
where
\be\beta_1=r_{H+}^2,~\beta_2=r_{H-}^2,~r_{H\pm}=\sqrt{\frac{l^2M}{2}
\left(1\pm\sqrt {1-\frac{4a^2}{l^2M^2}} \right)}\ee are the
horizons. In the following we assume both $\beta_{1,2}$ are real
meaning $\frac{4a^2}{l^2M^2}\leq1$. Obviously \be
\beta_1>\beta_2>0. \label{ccbeta} \ee We then transform the metric
to
\begin{eqnarray}
ds^2 & = & -\left(-M + \frac{r^2}{l^2} \right)dv^2+2dvdr-2adv d\hat \phi
+ r^2 d\hat\phi^2\label{Efc}\end{eqnarray}
by virtue of the change of variables
\begin{eqnarray}
dv& =&d t  - \frac{r^2l^2}{\cal K}dr,\label{v-t-r}\\
d\hat{\phi} & =&  d\phi - \frac{al^2}{ \cal
K}dr.\,\,\,\label{h-p-r}
\end{eqnarray}
Naturally $M,l,a$ are real and $M\geq 0,~l>0$. $a$ can be taken positive as
well since we always can account the change in the sign of $a$ by replacing
$\phi\to-\phi$ (or $\hat\phi\to-\hat\phi$)
The relations between the old and new coordinates are
\begin{equation}
v=v_0+t+\frac{l^2}{2(\beta_1-\beta_2)}\left(\sqrt{\beta_1}\log\frac{r-\sqrt
{
\beta_1}}{r+\sqrt{\beta_1}}-\sqrt{\beta_2}\log\frac{r-\sqrt{\beta_2}}{r+\sqrt{
\beta_2}}\right), \label{vtrans}
\end{equation}
\begin{equation}
\hat \phi=\hat
\phi_0+\phi+\frac{al^2}{2(\beta_1-\beta_2)}\left(\frac1{\sqrt{\beta_1}}\log\frac
{ r-\sqrt {
\beta_1}}{r+\sqrt{\beta_1}}-\frac1{\sqrt{\beta_2}}\log\frac{r-\sqrt{\beta_2}
} { r+\sqrt{ \beta_2}}\right),
\end{equation}
and we put $v_0=\hat\phi_0=0$. Note that $g_{00}$ component of
either metric vanishes only for \be r_H=l\sqrt{M}. \ee

{Clearly zero mass $M$ describes just an empty AdS
space. Thus the picture we have in mind is that the mass in fact
can be parametrized by step function $M\theta(v)$ where the parameter $M$ is 
constant, representing
the presence of the infalling null-shell at $v=0$, pure AdS
space-time for $v<0$, and BTZ BH solution for $v>0$. This is the
simplest form of the Vaidya solution which assumes a general
distribution of the mass $M(v)$.}

%%%%%%%%%%%%%%%%%%%%%%%%%%%%%%%%%%%%%%%%%%%%%%%%%%%%%%
%%%%%%%%%%%%%%%%%%%%%%%%%%%%%%%%%%%%%%%%%%%%%%%%%%%%%%

%\newpage
\section{Geodesics}
We are going to use the two point boundary correlation functions as
the thermalization probe. In $AdS/CFT$ they can be easily
approximated just by the renormalized length of a geodesic connecting
these two points on the boundary \cite{Balasubramanian:1999zv}:
\be \langle {\cal O}(t,\vec{x})) {\cal
O}(t^\prime,\vec{x}^\prime)\rangle_{\,_{ren}}\sim e^{-\Delta
L_{ren}}, \ee where $\Delta$ is the conformal dimension of the
operator $\cal O$ under consideration, and $L_{ren}$ is the
renormalized geodesic length.

So, once we are interested in the thermalization processes, we can
simply identify concepts of correlation functions and spacelike
geodesics ending up on the $AdS$ boundary. Since outside of the
thin shell the metric is indistinguishable from the one of a black
hole provided such a geodesic does not cross the shell, it has the same
length as it would have in the background of a black hole of mass
$M$, and therefore we may consider the corresponding correlator as
a thermalized one. If the geodesic crosses the shell, its length evolves
in time, and the corresponding correlator is out of the equilibrium.

\subsection{Proper time $\tau$ parametrization}

Equations for the geodesics in metric (\ref{BTZ-o}) are as follows
\begin{eqnarray}
\ddot r-\frac{{\cal K}}{r\,l^4} \dot{{t}} ^{2}+\frac{{\cal K}}{r\,l^2} \dot \phi ^{2}
-\frac{-r^4+a^2l^2}{r{\cal K}} \dot r ^{2}=0,\label{alex1}\\
\ddot{ t}-2\frac{r^3\,\dot{{t}}\dot r}{{\cal K}}+2\frac{a\,l^2r\dot\phi\dot r}{{\cal K}}
=0,\\
\ddot \phi-2\frac{a\,r\,\dot{{t}}\dot r}{{\cal
K}}+2\frac{(Ml^2-r^2)r\dot\phi\dot r}{{\cal K}} =0.\end{eqnarray}
Hereafter dot denotes a derivative with respect to the proper time
$\tau$. They can be obtained from the following corresponding
Lagrangian
\begin{eqnarray}\label{Lag}
{L} &=& -\frac12\left((-M+\frac{r^2}{ l^2}){\dot t}^2- \frac{\dot
r^2}{-M + \frac{r^2}{l^2} + \frac{a^2}{r^2}} +2a\dot{ t}\dot{\phi}
- r^2\dot{\phi}^2\right)\\\nn &=&
-\frac12\left((-M+\frac{r^2}{l^2}){\dot t}^2+ \frac{r^2l^2\dot
r^2}{{\cal K}} +2a\dot{ t}\dot{\phi} -
r^2\dot{\phi}^2\right),\end{eqnarray} which is $+1/2$ of the
metric. The conjugated momenta are
\begin{equation}
p_t=-(-M+\frac{r^2}{l^2}){\dot t}-a\,\dot{\phi},~
p_{{\phi}}=r^2\dot{\phi}-a\dot{ t},~ p_r= -\frac{r^2l^2\dot
r}{{\cal K}}.
\end{equation}
$\dot p_t=\dot p_{\phi}=0$ thanks to the fact that the initial metric does
not depend on $t$ and $\phi$ explicitly. Then we have
\begin{eqnarray}
{\cal E}&\equiv& p_t,\quad
{\cal J}\equiv p_{{\phi}},\\
{\dot t}&=&-\frac{{\cal E}r^2+{\cal
J}a}{(-M+\frac{r^2}{l^2})r^2+a^2}=\left({\cal E}r^2+{\cal
J}a\right)\frac{l^2}{{\cal K}},\\
\dot {\phi}&=&-\frac{{\cal E}a+{\cal
J}(M-\frac{r^2}{l^2})}{(-M+\frac{r^2}{ l^2})r^2+a^2}=\left({\cal
E}a+{\cal J}(M-\frac{r^2}{l^2})\right)\frac{l^2}{{\cal
K}}=\frac1{r^2}(a\dot t+{\cal J}).
\end{eqnarray}
%\newpage

Substituting these results into equation (\ref{alex1}) one obtains
\begin{equation}
\ddot r-\frac{{\cal K}}{r\,l^4} \dot{{t}} ^{2}+\frac{{\cal
K}}{r\,l^2} \dot \phi ^{2} -\frac{-r^4+a^2l^2}{r{\cal K}} \dot r
^{2}=\ddot r-A(r)\dot r ^{2}+B(r)=0, \label{metric1eqronly}
\end{equation}
where
\begin{equation*}
B(r)=\frac{1}{r{\cal K}}\left[l^2\left({\cal E}a+{\cal
J}(M-\frac{r^2}{ l^2})\right)^2-\left({\cal E}r^2+{\cal
J}a\right)^2\right],
\end{equation*}
and $A(r)$ is obvious. Than the latter equation can be recasted into a first
order linear differential equation for $X(r)=\dot r^2$ as follows
\begin{equation*}
X'-2A(r)X+2B(r)=0
\end{equation*}
with the general solution
\begin{equation*}
X(r)=\left(-2\int \left(Be^{-2\int
Adr}\right)dr+X_0/l^2\right)e^{2\int Adr}.
\end{equation*}
Computing everything we get
\begin{equation}
\dot r^2=\frac{1}{r^2l^2}\left[X_0{\cal K}+{\cal L}\right], \text{
where } {\cal L}={\cal J}^2Ml^2+2l^2{\cal E}{\cal J}a+r^2(l^2{\cal
E}^2-{\cal J}^2), \label{dr2r}
\end{equation}
computing the square root we come to two branches
\begin{equation}
\dot r=\pm\frac{1}{rl}\sqrt{X_0{\cal K}+{\cal L}}.
\label{dr2rsqrt}
\end{equation}
To simplify the succeeding analysis we introduce a few notations
\begin{equation}
\begin{split}
X_0{\cal K}+{\cal L}&=-X_0(r^2-\gamma_1)(r^2-\gamma_2),\\
\gamma_1+\gamma_2&=Ml^2+\frac{l^2{\cal E}^2-{\cal J}^2}{X_0},\\
\gamma_1\gamma_2&=l^2a^2-\frac{l^2{\cal J}}{X_0}(M{\cal J}+2a{\cal
E}).
\end{split}
\end{equation}

We note by product that (\ref{dr2rsqrt}) can be integrated out explicitly
\begin{equation}
\tau-\tau_0=
\frac{l}{2\sqrt{-X_0}}\log\left(r^2-\frac12(\gamma_1+\gamma_2)+\sqrt{(r^2-\gamma_1)(r^2-\gamma_2)}\right)
\label{dr2rint}
\end{equation}
and further inverted to give
\begin{equation}
r(y)=\frac1{\sqrt{2y}}\sqrt{\left(y+\frac{\gamma_1+\gamma_2}2\right)^2
-\gamma_1\gamma_2},\,\,\text{ where
}\,y=e^{2\frac{\sqrt{-X_0}}l(\tau-\tau_0)}. \label{dr2rsolrtau}
\end{equation}

Using equation (\ref{v-t-r}) one readily gets
\begin{equation}
\begin{split}
\dot v&=(r^2({\cal E}-\dot r)+a{\cal J})\frac{l^2}{\cal K}.
\end{split}
\label{vdot}
\end{equation}
This expression is important since it determines whether the geodesics are
thermal or not as
we shall see shortly.

\subsection{Radial coordinate $r$ parametrization}

Hereafter we introduce the prime for a derivative with respect to $r$.
Computing $t(r)$ we write
\begin{equation}
t'=\frac{\dot t}{\dot r}=\frac{\left({\cal E}r^2+{\cal
J}a\right)\frac{l^2}{{\cal K}}}{\pm\frac{1}{rl}\sqrt{X_0{\cal
K}+{\cal L}}}\,.
\end{equation}
To find out $t(r)$ we have to integrate the RHS with respect to $r$.
Therefore the integral of interest is
\begin{equation}
I=\int\frac{x-\alpha}{(x-\beta_1)(x-\beta_2)\sqrt{(x-\gamma_1)(x-\gamma_2)}}dx,
\end{equation}
where $x=r^2$.
This can be computed with the following result:
\begin{equation}
\begin{split}
I=\frac1{(\beta_1-\beta_2)}\bigg[&\frac{\alpha-\beta_1}{\sqrt{B_1}}\ln(
X_1
-\sign(x-\beta_1)\sqrt{X_1^2-(\gamma_1-\gamma_2)^2})-\\
-&\frac{\alpha-\beta_2}{\sqrt{B_2}}\ln( X_2
-\sign(x-\beta_2)\sqrt{X_2^2-(\gamma_1-\gamma_2)^2})\bigg]\equiv
I_-,
\end{split}
\label{mainIanswer}
\end{equation}
where
$$
B_{1,2}=(\beta_{1,2}-\gamma_2)(\beta_{1,2}-\gamma_1),\quad
X_{1,2}=-(2\beta_{1,2}-\gamma_1-\gamma_2)-\frac{
2B_{1,2}}{x-\beta_{1,2}}.
$$
An important technical note here is that accounting for two branches we must
have
$\pm I_-$ and we note that
\begin{equation}
\begin{split}
-I_-=&\frac1{(\beta_1-\beta_2)}\bigg[\frac{\alpha-\beta_1}{\sqrt{B_1}}\ln(
X_1
+\sign(x-\beta_1)\sqrt{X_1^2-(\gamma_1-\gamma_2)^2})-\\
-&\frac{\alpha-\beta_2}{\sqrt{B_2}}\ln( X_2
+\sign(x-\beta_2)\sqrt{X_2^2-(\gamma_1-\gamma_2)^2})\bigg]+C_\alpha
=I_++C_\alpha,
\end{split}
\end{equation}
where the signs in front of $\sign(x-\beta_{1,2})$ inside the logs are altered.
The constant $C_\alpha$ is given by
\begin{equation*}
\begin{split}
C_\alpha=-\frac2{(\beta_1-\beta_2)}\bigg[\frac{\alpha-\beta_1}{\sqrt{B_1}}
-\frac{\alpha-\beta_2}{\sqrt{B_2}}\bigg]\ln(\gamma_1-\gamma_2).
\end{split}
\end{equation*}
In our
study we have to take different branches based on the sign in front of this
square root making the second form more convenient to us. The first form with
an overall minus sign is still valid but would force us having different
integration constants to glue both branches together. Surely, $-I_+\neq I_-$.

Thus we have
\begin{equation}
\begin{split}
t(r)=t_0-\frac{{\cal
E}l^3}{2\sqrt{-X_0}}{I_\mp}|_{\alpha=-\frac{{\cal J}a}{\cal E}}\,
.\end{split}\label{trfull}
\end{equation}
{Calculations for $\phi$ are identical just with the different
parameter $\alpha$:}
\begin{equation}
\phi'=\frac{\dot\phi}{\dot r}=\frac{\left({\cal E}a+{\cal
J}(M-\frac{r^2} {l^2})\right)\frac{l^2}{{\cal
K}}}{\pm\frac{1}{rl}\sqrt{X_0{\cal K}+{\cal L}}},
\end{equation}
providing
\begin{equation}
\begin{split}
\phi(r)=\phi_0+\frac{{\cal
J}l}{2\sqrt{-X_0}}{I_{\mp}}|_{\alpha=\frac{{\cal E}a+{\cal
J}M}{{\cal J}/l^2}}.\end{split}\label{phirfull}
\end{equation}
Some nice cancelations happen here. Namely \be
\begin{split}
\frac{\alpha-\beta_{1,2}}{\sqrt{B_{1,2}}}|_{\alpha=-aJ/{\cal
E}}=-s_{1,2}\frac{\sqrt{\beta_{1,2}}}{l\cal E},\quad
\frac{\alpha-\beta_{1,2}}{\sqrt{B_{1,2}}}|_{\alpha=Ml^2+l^2a{\cal
E}/{\cal J}}=s_{1,2}\frac{\sqrt{\beta_{2,1}}}{J},
\end{split}
\label{magic}
\ee
where
\be
\begin{split}
B_1&=(\sqrt{\beta_1}{\cal E} l+\sqrt{\beta_2}{\cal J})^2=S_1^2,\quad
B_2=(\sqrt{\beta_2}{\cal E} l+\sqrt{\beta_1}{\cal J})^2=S_2^2,\\
s_1&=\sign(S_1),~s_2=\sign(S_2).
\end{split}
\label{newBB}
\ee
Hence finally we get
\begin{equation}
\begin{split}
t(r)=t_0+\frac{l^2}{2\sqrt{-X_0}(\beta_1-\beta_2)}\left(s_1\sqrt{\beta_1}\ln
F_{1\mp}-s_2\sqrt{\beta_2}\ln F_{2\mp}\right),
\end{split}\label{trfullmagic}
\end{equation}
\begin{equation}
\begin{split}
\phi(r)=\phi_0+\frac{l}{2\sqrt{-X_0}(\beta_1-\beta_2)}\left(s_1\sqrt{
\beta_2}\ln F_{1\mp}-s_2\sqrt{\beta_1}\ln
F_{2\mp}\right),\end{split}\label{phirfullmagic}
\end{equation}
\begin{equation*}
F_{1,2\mp}=X_{1,2}\mp\sign(x-\beta_{1,2})\sqrt{X_{1,2}^2-(\gamma_1-\gamma_2)^2},
\quad x=r^2.
\end{equation*}

%%%%%%%%%%%%%%%%%%%%%%%%%%%%%%%%%%%%%%%%%%%%%%%%%%%

\subsection{Geodesics of interest which start and finish at $r=\infty$}

Here we specialize to the type of geodesics which begin and end at
the boundary $r=\infty$. Returning to (\ref{dr2rsqrt}) we see that
for geodesics reaching the boundary $r=+\infty$ we should have
$X_0=-1$. Moreover, we must guarantee that a turning point exists
meaning that at least one of $\gamma_{1,2}>0$. If both are greater
than $0$ we assume \be\gamma_1>\gamma_2.\label{ccgamma}\ee Also we
assume the geodesics of interest do not cross any of the horizons.
Writing down some elementary school level formulae we see that
assuming $\gamma_1+\gamma_2=2b, \gamma_1\gamma_2=c$ one has
$$
\gamma_{1,2}=b\pm\sqrt{b^2-c}.
$$
Now there are two cases
\begin{itemize}
  \item $c>0$\\
  In this case we must worry about $b^2>c$ so that roots exist an then $b$ must
be positive. Indeed, if it is negative then both roots are negative as well. If
all the conditions are met the greater root is given by the plus sign
  \item $c<0$\\
  In this case we must not worry about $b^2>c$ as well as $b$ can be any. The
greater root is given by the plus sign as in the previous case.
\end{itemize}
Hence the turning point we are interested in is
\begin{equation}
r^2=\gamma_1=b+\sqrt{b^2-c}\,\text{ with }\left\{
\begin{tabular}{l}
$c<0$ or\\
$c>0,~b>0$ and $b^2>c$.
\end{tabular}
\right.\label{degree2}
\end{equation}
%At the turning point we have
%\begin{equation*}
%F_{1,2\mp}=X_{1,2}
%\end{equation*}
Relations (\ref{newBB}) imply $B_{1,2}>0$ which in turn implies
$\beta_1<\gamma_2$ and one can check that this leaves only with the second
possibility in (\ref{degree2})

\begin{figure}[!h]
\centering
\includegraphics[width=33mm]{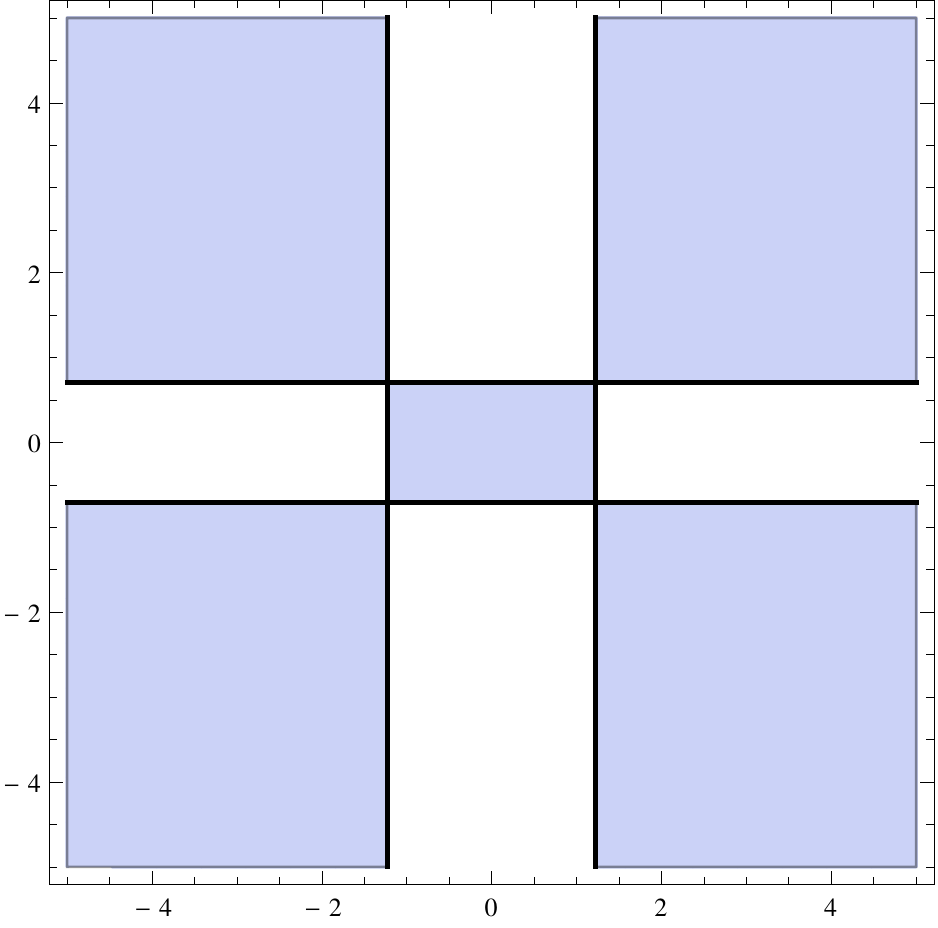}a
\includegraphics[width=33mm]{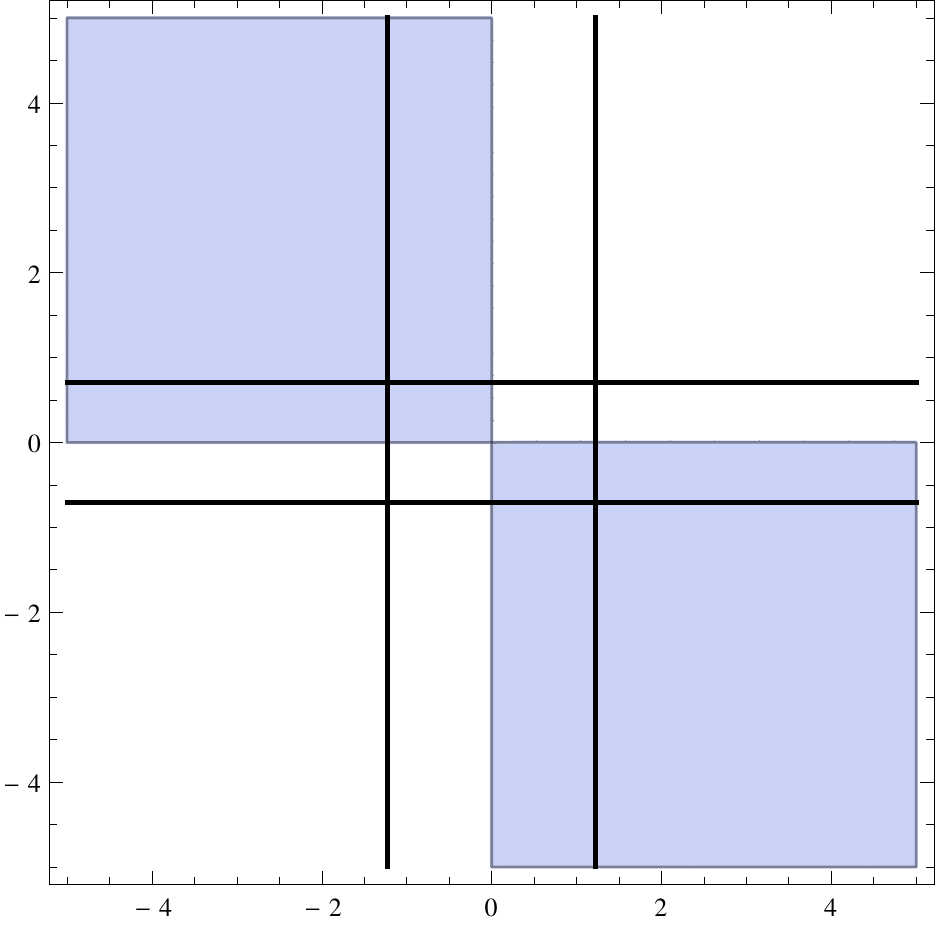}b
\includegraphics[width=33mm]{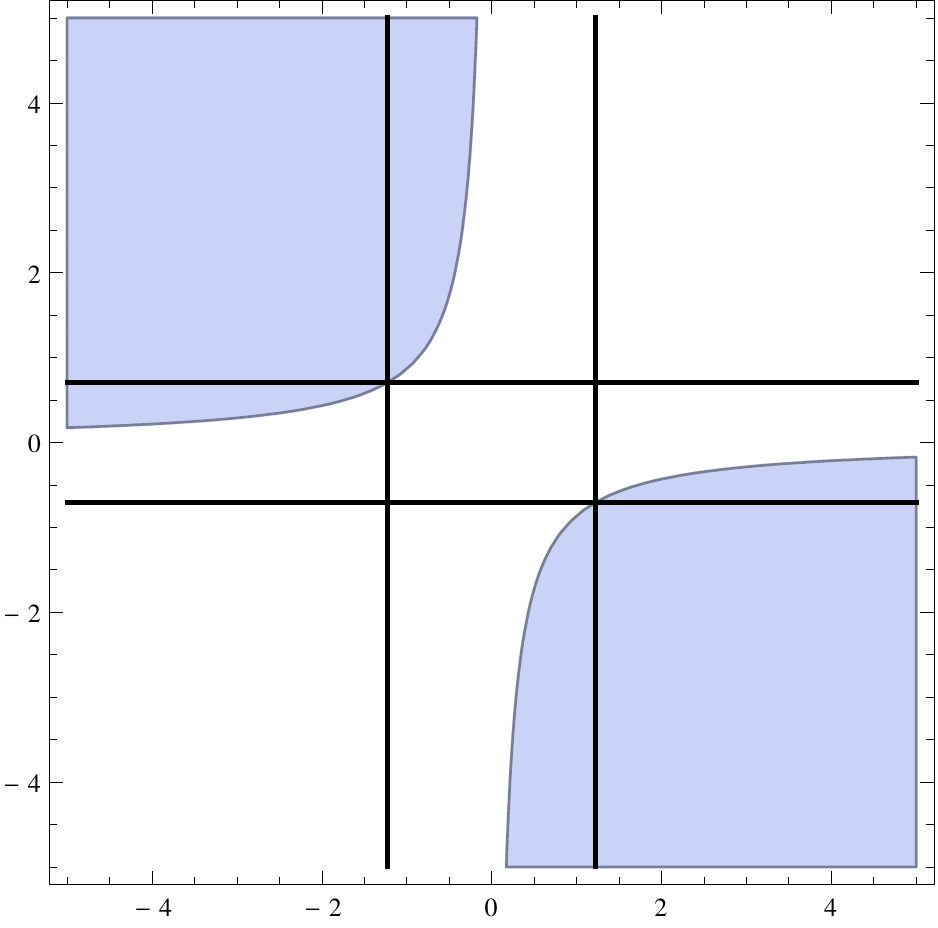}c
\includegraphics[width=33mm]{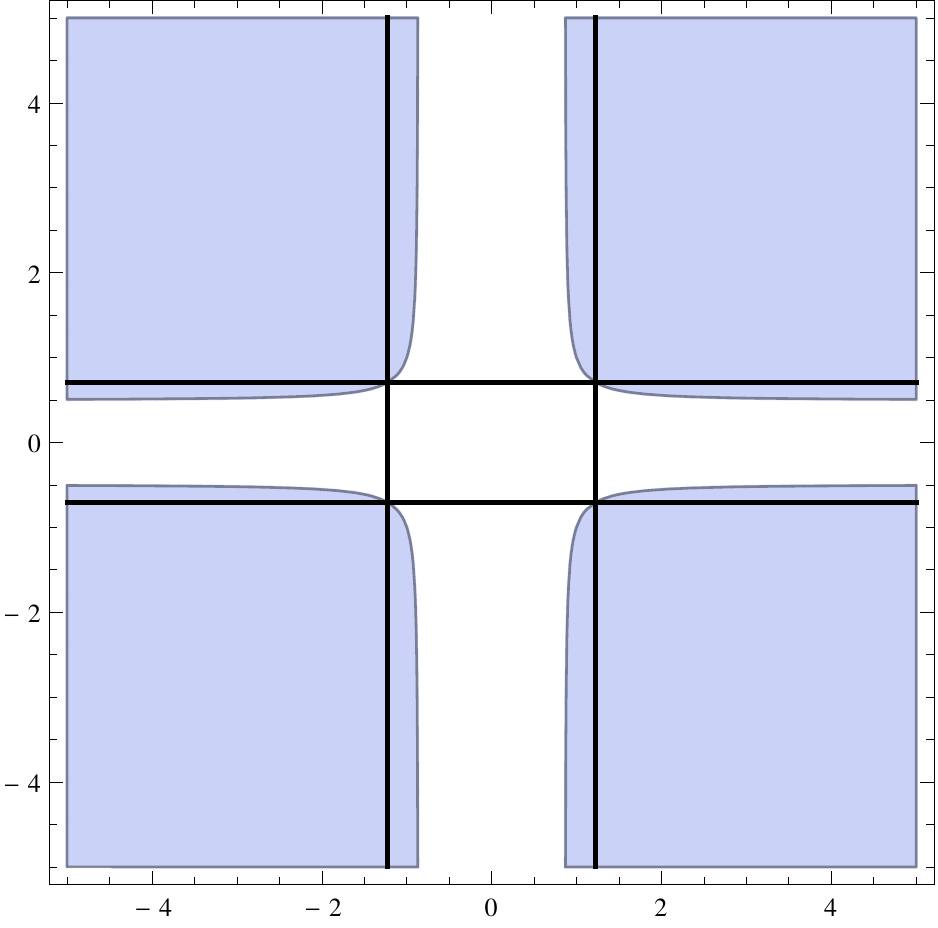}d\\
$~$\\
\includegraphics[width=33mm]{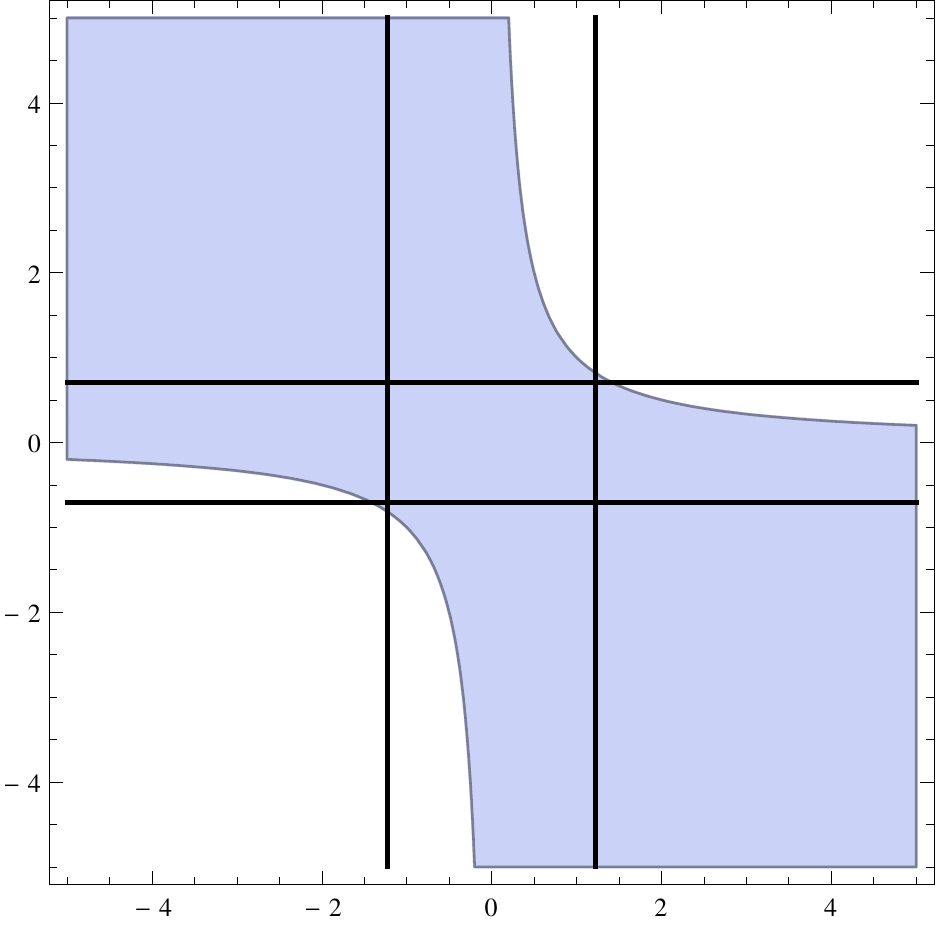}e
\includegraphics[width=33mm]{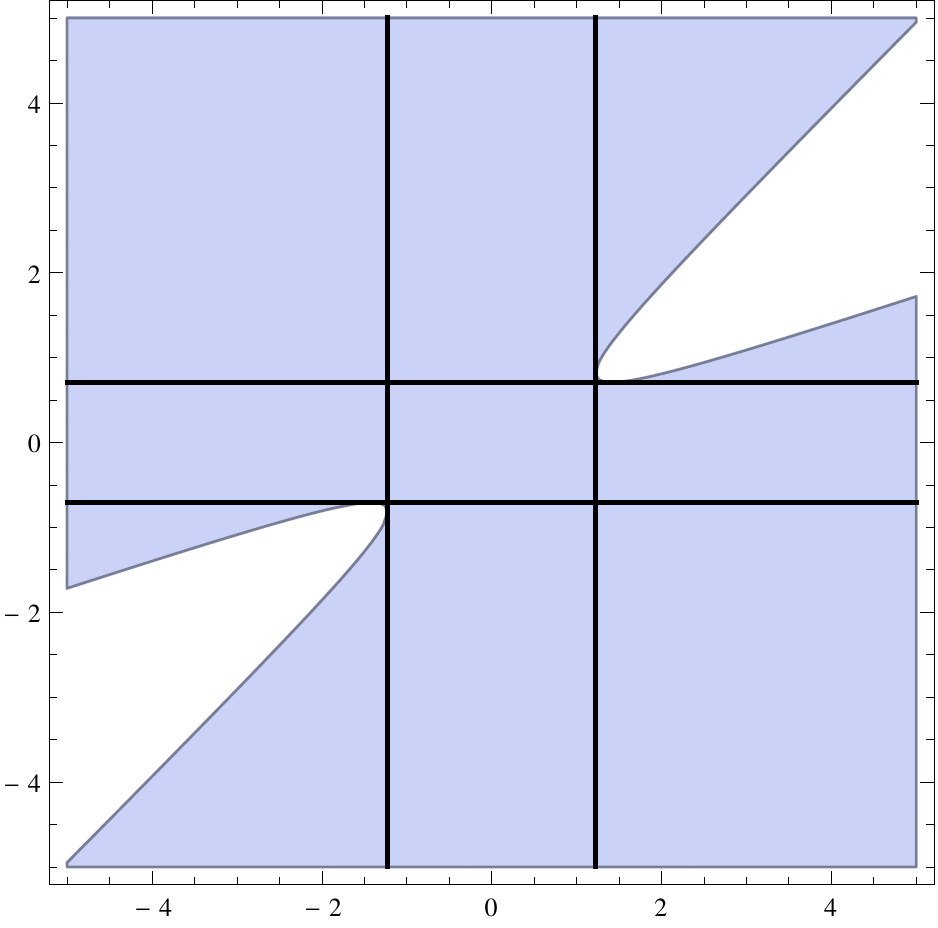}f
\includegraphics[width=33mm]{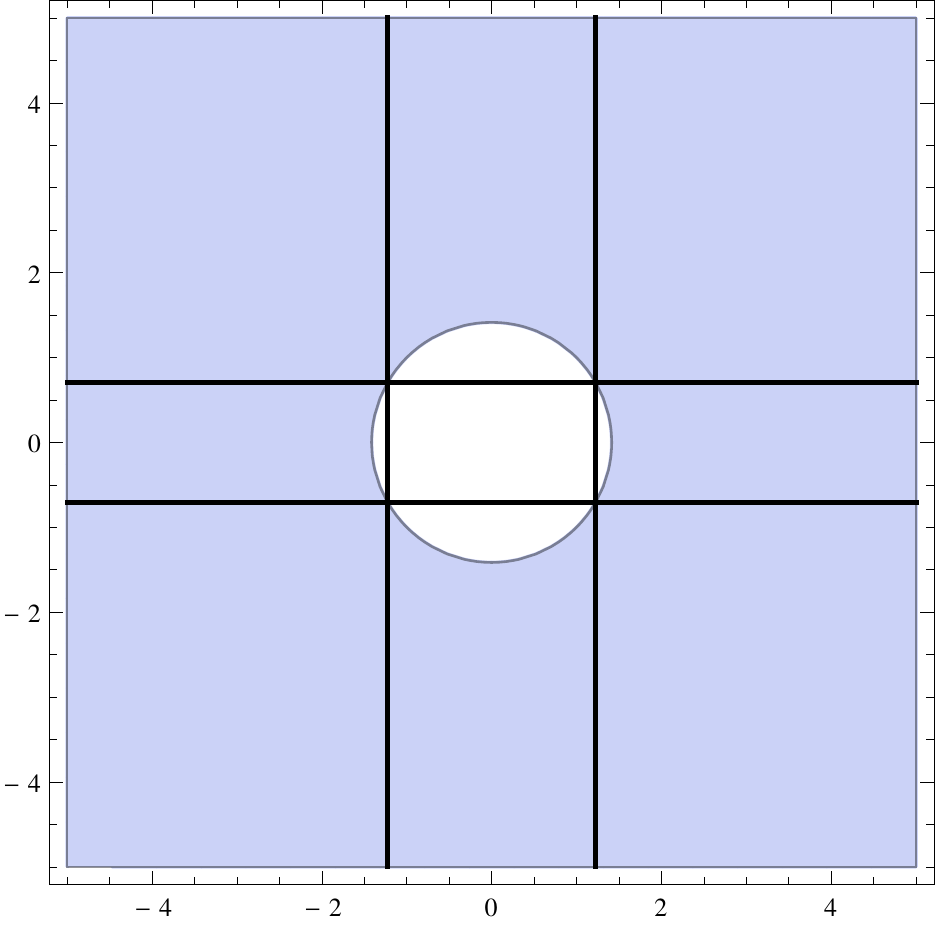}g
\includegraphics[width=33mm]{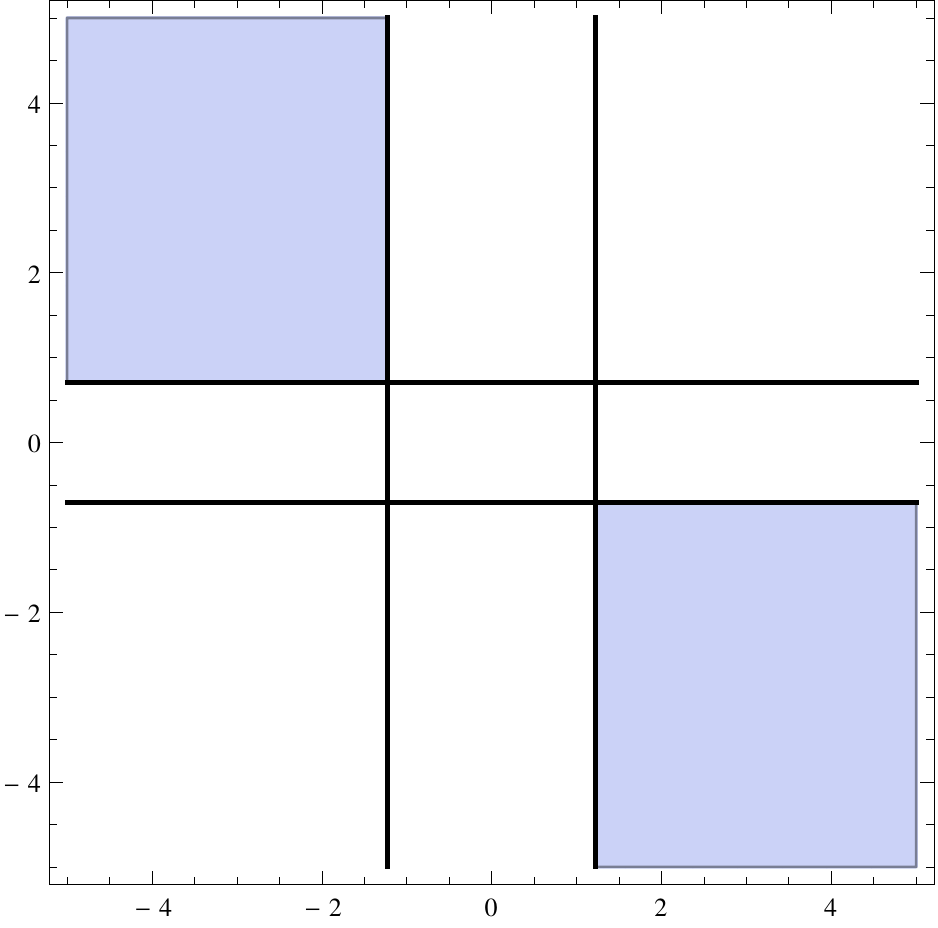}h
\caption{Domains for $M=1,~l=1,~a=1/4$. Horizontal axis is $\rho_-$ and
vertical one is $\rho_+$, vertical black lines are
$|\rho_-|=\sqrt{l}\mu_1=\sqrt{3/2}$, horizontal black lines are
$|\rho_+|=\sqrt{l}\mu_2=\sqrt{1/2}$.}
\label{domains}
\end{figure}

So in total we have \be \beta_2<\beta_1<\gamma_2<\gamma_1,
\label{cccc} \ee \be \gamma_1+\gamma_2>0,\gamma_1\gamma_2>0,
\label{cccc2} \ee \be
b^2-c=\frac{(\gamma_1+\gamma_2)^2}4-\gamma_1\gamma_2=\frac{(\gamma_1-\gamma_2)^2
}4>0.\label{cccc22} \ee The last condition (\ref{cccc22}) is not
trivial from the very beginning once we go back to the parameters
of our problem. It is read \be (Ml^2-l^2{\cal
E}^2+J^2)^2-4(l^2a^2+{l^2J}(MJ+2a{\cal E}))=(J^2+l^2({\cal
E}^2-M))^2-4l^2({\cal E}J+a)^2>0. \label{cccctrivial} \ee We
rewrite it as follows \be
(\rho_-+\sqrt{l}\mu_1)(\rho_--\sqrt{l}\mu_1)(\rho_++\sqrt{l}
\mu_2)(\rho_+-\sqrt{l}\mu_2)>0,
\ee where \be\label{mu}
\mu_1=\sqrt{Ml+2a}>0,\quad\mu_2=\sqrt{Ml-2a}>0,\quad \rho_+={\cal
E}l+J,\quad\rho_-={\cal E}l-J. \ee Now we can readily see the
solution \bea &&|\rho_-|>\sqrt{l}\mu_1,|\rho_+|>\sqrt{l}\mu_2\,
\text{ or
}\label{outer}\\
&&|\rho_-|<\sqrt{l}\mu_1,|\rho_+|<\sqrt{l}\mu_2\label{inner} \eea
which are five domains on $(\rho_-,\rho_+)$ plane
(Fig.~\ref{domains}a). Horizons can be written in terms of
$\mu_{1,2}$ as follow \be \label{beta} \beta_{1}=\frac l4(\mu_1+
\mu_2)^2,\quad \beta_{2}=\frac l4(\mu_1- \mu_2)^2. \ee Note that
due to $a>0$ condition we have $\mu_1>\mu_2$.

Condition (\ref{cccc}) can be taken as $\beta_1+\beta_2<\gamma_1+\gamma_2$
which is explicitly
\begin{equation}
\begin{split}
\rho_-\rho_+<0.
\end{split}
\label{ccEJrhopm}
\end{equation}
This is Fig.~\ref{domains}b.

The less simple relation is $\beta_1<\gamma_2$ is hold upon: \be
\rho_-\rho_+<-l\mu_1\mu_2, \label{ccsuperceed} \ee and \be
\frac{\mu_1^2}{\rho_-^2}+\frac{\mu_2^2}{\rho_+^2}<\frac 2l \ee
(\ref{ccsuperceed}) is depicted on Fig.~\ref{domains}c, and this
supersedes (\ref{ccEJrhopm}). The last inequality holds in the
domain (\ref{outer}), see Fig.~\ref{domains}d.

Examining (\ref{cccc2}) we write \be Ml^2>\rho_-\rho_+ \ee which
is true thanks to (\ref{ccsuperceed}) (see Fig.~\ref{domains}e),
and \be
\mu_1^2\rho_+^2+\mu_2^2\rho_-^2-(\mu_1^2+\mu_2^2)\rho_+\rho_-+\frac
l4(\mu_1^2-\mu_2^2)^2>0, \ee and this clearly holds in the domain
(\ref{ccEJrhopm}) which is enough for our purposes (see also
Fig.~\ref{domains}f).

The intersection of all the conditions is depicted graphically in
Fig.~\ref{domains}h.

\section{Thermalization}

Thermal geodesics are those inside the shell and we thus must find out whether
geodesics cross the shell or not.
This question is translated into the analysis of $\dot v=0$ point using
(\ref{vdot}). This happens when
\begin{equation}
  r^2({\cal E}-\dot r)+a{\cal J}=r^2{\cal E}+a{\cal J}\mp\frac
rl\sqrt{(r^2-\gamma_1)(r^2-\gamma_2)}=0. \label{star1}
\end{equation}
Here we have taken already
$X_0=-1$. Then we come to the algebraic condition
\begin{equation}
l^2(r^2{\cal E}+a{\cal J})^2=r^2(r^2-\gamma_1)(r^2-\gamma_2)
\end{equation}
which can be explicitly solved for $r^2$ as
\begin{equation}
r^2\text{ is equal to }{\cal J}^2,\beta_1,\text{ or }\beta_2.
\end{equation}
We assume $\dot v=0$ point does not coincide with a horizon and
analyze $r=|{\cal J}|$ then assuming moreover ${\cal J}^2>\beta_1$ (to avoid
the horizon
crossing).
Also as mentioned before the turning point is $\gamma_1$ and in order to have
the thermalization we must require ${\cal J}^2\geq \gamma_1$,
which is always true upon
\be
Ml^2<l^2{\cal E}^2+{\cal J}^2
\ee
to be one extra condition in the game. It is however always satisfied in the
domain
(\ref{outer}), see also Fig.~\ref{domains}g.

All the quantities associated with the $\dot v=0$ point hold the
star subscript. Our goal now is to compute $v_*$. First we define
\be r_*=|{\cal J}| \ee and introduce \be
\chi=(x-\gamma_1)(x-\gamma_2). \ee Note that at the $\dot v=0$
point $\chi=l^2({\cal EJ}+a)^2$. Equation (\ref{star1}) evaluated
at $r=r_*$ reads
\begin{equation}
r_*^2{\cal E}+a{\cal J}\mp
{r_*}|{\cal EJ}+a|=0
\label{starbranch}
\end{equation}
or a bit more complicated
\begin{equation}
{\cal J}({\cal EJ}+a)[1\mp s_{\cal J}s]=0,\text{ where }s_{\cal
J}=\sign({\cal J}),~s=\sign({\cal EJ}+a). \label{starbranchsigns}
\end{equation}
\begin{itemize}
  \item
If both signs are the same then we get the upper sign here.
\item
If the signs are opposite then we get the lower
sign.
\end{itemize}
All the other formulae must be taken with the appropriate sign respectively.

Several quantities are important to be introduced \be
t_{\mp}=t_{\mp}(\infty),\quad\phi_{\mp}=\phi_{\mp}(\infty),\quad
\Delta\phi=\phi_+-\phi_-. \ee Then with the help of intermediate
calculations given in the Appendix one yields
\begin{equation}
\begin{split}
v_*=& \frac{t_++t_-}2+\sign({\cal J})l\frac{\Delta\phi}2.
\end{split}
\label{kerrnokerr}
\end{equation}
Since $v_*$ is the minimal value of the $v$ coordinate on the geodesics the
condition $v_*>0$ guarantees the corresponding geodesics is always under the
shell. This in turn is the sufficient condition for the thermalization to
happen.

Surprisingly, this formula is the same as for $a=0$ case. No
question that the space separation $\Delta\Phi$ is different but
the structure is identical. Consequently the thermalization time
as a function of the space separation is the same. It is just
equal to the probe length $\Delta\Phi$. This makes us thinking
that other results like non-thermal geodesics analysis or analysis
of geodesics which cross horizons can be easily or perhaps even
identically generalized to the Kerr-BTZ space-time.

%%%%%%%%%%%%%%%%%%%%%%%%%%%%%%%%%%%%%%%%%%%%%%%%%%%%%%%%%%%%%%%%%%
%%%%%%%%%%%%%%%%%%%%%%%%%%%%%%%%%%%%%%%%%%%%%%%%%%%%%%%%%%%%%%%%%%

\section{Conclusions}

In this paper we have considered thermalization of a particular
kind of non-local observables, two-point non-equal time
correlation functions, in (1+1)-dimensional quantum field theory dual to the
AdS-BTZ-Vaidya space-time. Our results are in agreement with those
of \cite{Lopez}, moreover, we have shown that non-zero angular
momentum does not affect the formula which selects thermal geodesics at
all.

However many other issues deserve more detailed analysis. In particular
if one is talking about bulk/boundary duality he might wonder
whether bulk region bounded by a minimal surface as introduced in
\cite{Ryu-Takayanagi} is really the most natural dual to the
boundary region ${\cal A}$. This question is still under
discussion, and an alternative object
named causal wedge (and its boundary dual --- the holographic causal
information) has been defined in \cite{Hubeny-Rangamani}. Without getting into
details we just
mention that this wedge is a region in the bulk which can be
causally defined by boundary conditions on ${\cal A}$.

Non-trivial issue is arising when we are interested in
out-of-equilibrium dynamics in AdS/CFT. Suppose we consider a time
dependent bulk solution and a dual non-equilibrium field theory, is
the causality preserved in such a system or not? In other words, is it
true that throughout the whole evolution the boundary region and
its bulk counterpart are casually related?
This question was addressed in \cite{Hubeny} for a particular case
of non-rotating AdS-Vaidya metric, and this test of time evolution
of holographic causal information has shown that this observable
is evolving in a normal causal way.
The question how the conclusions might change provided we consider the
Kerr-BTZ space-time, which causal structure is very different due
to the presence of two event horizons is still open.

Another problem to be analyzed is the possible relation between
AdS-Vaidya space-times and the HIC phenomenology. Although here
we are discussing only $AdS_3/CFT_2$ case, we
may interpret the AdS-Kerr case as a dual to HIC with non-zero
impact parameter in higher dimensions. {There are two possibilities how the
centrality dependence of
the thermalization process quantities may appear. The first one is related
with the geometry of the dual space, i.e. with the fact that we deal with
Kerr-AdS, and the second one is just based on a simple collision
picture of two ``pancakes'', since now the maximal distance between
points in cross-section area is $2\sqrt{R^2_A-b^2}$ instead of
$2R_A$ where $R_A$ is the radius of the ion, and $b$ is the impact
parameter of the collision. The second scenario does not depend on
the dimension of the space-time, while the first one can.}

\acknowledgments
AK is grateful to Ben Craps for valuable discussions.

Authors are supported in part by the RFBR grant 11-01-00894. A.B.
is supported by the Dutch Foundation for Fundamental Research on
Matter (FOM) and RFBR grant 12-01-31298. A.K. is supported by an
``FWO-Vlaanderen'' postdoctoral fellowship and also supported in
part by Belgian Federal Science Policy Office through the
Interuniversity Attraction Pole P7/37, the ``FWO-Vlaanderen''
through the project G.0114.10N.

%%%%%%%%%%%%%%%%%%%%%%%%%%%%%%%%%%%%%%%%%%%%%%%%%%%%%%%%%%%%%%%%%%
\appendix
\section{Derivation of (\ref{kerrnokerr})}

First we massage the logarithms in integration result (\ref{mainIanswer})
as
follows
{
\begin{equation*}
\begin{split}
F_{1,2\mp}=\frac{
(x-\beta_{1,2})^2-\chi-B_{1,2}\mp2\sqrt{B_{1,2}\chi}}{
x-\beta_{1,2}},
\end{split}
\end{equation*}
}
where we dropped $\sign(x-\beta_{1,2})$ factor as it is equal to $+1$ here.
Then from (\ref{vtrans}) one has
\begin{equation}
v_*=t_*+\frac{l^2}{2(\beta_1-\beta_2)}\left(\sqrt{\beta_1}\log\frac{
r_*-\sqrt { \beta_1}}{r_*+\sqrt {
\beta_1}}-\sqrt{\beta_2}\log\frac{r_*-\sqrt{\beta_2}}{r_*+\sqrt{\beta_2}}\right),
\label{prevstar}
\end{equation}
\begin{equation}
\begin{split}
t_*=&\frac{l^2}{2(\beta_1-\beta_2)}\bigg[s_1\sqrt{\beta_1}
\ln\left(
{\frac{({\cal J}^2-\beta_1)^2-l^2({\cal
E}{\cal J}+a)^2-B_1-2ls_1s_{\cal J}({
\cal E}{\cal J}+a){S_1}}{{\cal J}^2-\beta_1}}\right)-\\
-&s_2\sqrt{\beta_2}\ln\left( \frac{({\cal
J}^2-\beta_2)^2-l^2({\cal E}{\cal J}+a)^2-B_2-2ls_2s_{\cal
J}({\cal E}{\cal J}+a){S_2}}{{\cal J}^2-\beta_2}\right)\bigg]+t_0,
\end{split}
\end{equation}
where we took $\mp s=-s_{\cal J}$ from (\ref{starbranchsigns}).
Factorizing, simplifying, and regrouping terms a number of times
one can get
\begin{equation}
\label{prevstar3}
\begin{split}
v_*=&
\frac{t_++t_-}2-\frac{l^2}{2(\beta_1-\beta_2)}
\bigg[s_1\sqrt{\beta_1}-s_2\sqrt{\beta_2}\bigg]
\ln(\gamma_1-\gamma_2)+\\
+&\frac{l^2}{2(\beta_1-\beta_2)}\bigg[{s_1\sqrt{\beta_1} }
\ln[{\cal J}^2-{\cal E}^2l^2+\beta_1-\beta_2-2s_1s_{\cal J}S_2]-\\
&\qquad\qquad- {s_2\sqrt{\beta_2}}\ln[{\cal J}^2-{\cal
E}^2l^2+\beta_2-\beta_1-2s_2s_{\cal J}S_1]\bigg],
\end{split}
\end{equation}
where we have used \be
t_0=\frac{t_++t_-}2-\frac{l^2}{2(\beta_1-\beta_2)}
\bigg[s_1\sqrt{\beta_1}-s_2\sqrt{\beta_2}\bigg]
\ln(\gamma_1-\gamma_2). \ee

Now we compute explicitly $\Delta\phi$.
We transform it to look similar to (\ref{prevstar3}) as
follows
\be
\label{predphi2}
\begin{split}
\Delta
\phi=&-\frac{l}{(\beta_1-\beta_2)}\bigg[
{s_1s_{\cal J}\sqrt{\beta_1}}-{s_2s_{\cal J}\sqrt{\beta_2}}
\bigg]
\ln(\gamma_1-\gamma_2)+\\
+&\frac{l}{(\beta_1-\beta_2)}\bigg[
{s_1s_{\cal J}\sqrt{\beta_1}}\ln\left(
{{\cal J}^2-l^2{\cal
E}^2+\beta_{1}-\beta_{2}-s_1s_{\cal J}2S_{2}}\right)-\\
&\qquad\qquad-{s_2s_{\cal J}\sqrt{\beta_2}} \ln\left( {{\cal
J}^2-l^2{\cal E}^2+\beta_{2}-\beta_{1}-s_2s_{\cal
J}2S_{1}}\right)\bigg].
\end{split}
\ee Juxtaposing (\ref{prevstar3}), and (\ref{predphi2}) it becomes
clear that we come to (\ref{kerrnokerr}).
%%%%%%%%%%%%%%%%%%%%%%%%%%%%%%%%%%%%%%%%%%%%%%%%%%%%%%%%%%%%%%%%%%

%\input{Kerr-AdS-bib-IA.tex}

\end{document}